\documentclass[twocolumn,showpacs,preprintnumbers,amsmath,amssymb]{revtex4}
\usepackage{graphicx}% Include figure files
\usepackage{dcolumn}% Align table columns on decimal point
\usepackage{bm}% bold math

\begin{document}

\title{Comment on ``An Improved Experimental Limit on the Electric-Dipole Moment
of the Neutron," C.A. Baker, et al. \cite{1}}

\author{S.K. Lamoreaux}\altaffiliation{email address: steve.lamoreaux@yale.edu}
\affiliation{Yale University, Department of Physics, Box 208120,
New Haven, CT 06520-8120}

\author{R. Golub}\affiliation{North Carolina State University, Department of Physics,
Campus Box 8202, Raleigh, NC 27695-8202}

\date{Sept. 28, 2006}

\begin{abstract}

\end{abstract}
 \maketitle

In a recent measurement of the the neutron electric dipole moment
(EDM) \cite{1}, a new systematic ``geometric phase" effect was
discovered. To
account and correct for this effect, a masterful series of
measurements and analysis were implemented. We would like to point
out an effect that was overlooked in the analysis.
Because the
measuring apparatus is on the Earth, it is in a rotating coordinate system, 
so there will be an apparent Larmor frequency shift
that depends on both the projection of the quantization axis $\hat B$
(magnetic field) on the Earth's angular rotation axis $\hat\Omega$, and on the
relative direction of Larmor precession which depends on the sign of the
magnetic moment. This additional rotation creates a change in the
apparent magnetic moment, and is of opposite sign for magnetic
field ``up" vs. ``down."  This effect has been used to construct nuclear spin
gyroscopes.\cite{3} 

Normally this would not affect the EDM measurement as the shift is
independent of electric field. However the authors use the
deviation of the ratio of neutron and Hg co-magnetometer
frequencies, $|\omega_{n}/\omega_{Hg}|$ from its expected value,
$|\gamma_{n}/\gamma_{Hg}|$, as a measure of the volume averaged
magnetic field gradient, $\left\langle \partial B_{z}/\partial
z\right\rangle _{V}\equiv G$, in the apparatus. The two quantities
are related because there is a small offset in the centers of mass
of the two gases caused by gravity. The gradient determined in
this way is then used to calculate the systematic contribution to
the measured EDM due to the ``geometric phase" \cite{pen,lamgol}
effect. In order to establish a zero for this effect the authors
then make use of the fact that the slope of the relation between the gradient
and the frequency ratio deviation changes sign when the direction
of magnetic field is reversed. Then the shift in resonance frequency 
of the two species due to the Earth's
rotation alters the gradient vs. frequency ratio relationship in such a
way as to mimic a true EDM.  The experiment is sensitive to a correlation of
the form $\hat B\cdot\hat\Omega$;  a true time reversed state requires changing the sign of $\hat\Omega$
in addition to the other reversals used to discriminate an EDM.

In \cite{1}, the ratio of Larmor frequencies is compared to the ratio of gyromagnetic ratios,
\[
R_a=\left|  \left(  \omega_{n}/\omega_{Hg}\right)  \left(  \gamma_{Hg}%
/\gamma_{n}\right)  \right|.
\]
Thus in the presence of the Earth's rotation the relation between the
frequency ratio deviation and $G$ is given by (noting that $\gamma_n<0$, $\gamma_{Hg}>0$)%
\begin{equation}\label{4}
\frac{G|\Delta h|}{B_{0}} =\pm  \left(  R_a-1\right)  +\left( \frac{\omega_{\oplus}\sin\theta_L}{B_{0}}\right)\left( \frac{1}
{\gamma^{\prime}}\right)
\end{equation}
where $\Delta h$ is the offset in centers of mass, the plus sign
is for $B_{0}$ pointing down,  $\omega_\oplus/2\pi=11.6\ \mu$Hz,
the sidereal rotation frequency of the Earth, and $\theta_L\approx
45^\circ$ is the latitude of the experiment's location (Grenoble.
France), and ${(\gamma^{\prime})}^{-1}  =|\gamma_{n}|^{-1}+|\gamma_{Hg}
|^{-1}$

The authors used the observation of the post-storage ultracold neutron polarization to determine the
value of $\left(  R_a-1\right)  $ corresponding to zero gradient so that from
Eq. (\ref{4}) (neglecting other differential effects, listed in Table I of \cite{1})%
\begin{equation}
\left(  R_{a0}-1\right)  =\mp\left( \frac{\omega_{\oplus}\sin\theta_L}{B_{0}}\right)\left( \frac{1}
{\gamma^{\prime}}\right).
\end{equation}
The authors state, ``There are some processes that can interfere with the above
GP error removal - essentially any process that changes $R_{a}$ and/or
$d_{n,Hg,f}$ without conforming to the ratio between the two given by Eq. (4) [of \cite{1}]
and where, in addition, the changes differ with the direction of
\textbf{B}$_{0}$." The effect of the Earth's rotation satisfies these
criteria.
The authors also state that, ``The crossing point $\left(  R_{a0},d_{n}^{\prime
}\right)  $ [of the two lines defined by their Eq. (5)] provides an
estimate of $d_{n}^{\prime}$ free of $d_{n,Hg,f}$." (The latter symbol
represents the correction to the neutron edm due to the geometric phase
frequency shift in $^{199}$Hg.) However it is easy to see that the crossing point
occurs at
\begin{align}
R^{\ast} &  =\frac{R_{a0\downarrow}+R_{a0\uparrow}}{2}\label{5}\\
d_{meas} &  =d_{n}^{\prime}+k\left[  \frac{\left(  R_{a0\uparrow}%
-R_{a0\downarrow}\right)  }{2}\right]  \\
&  =d_{n}^{\prime}+k\left[  \left( \frac{\omega_{\oplus}\sin\theta_L}{B_{0}}\right)\left( \frac{1}
{\gamma^{\prime}}\right)  \right]  \label{6a}%
\end{align}
where the up/down arrows refer to magnetic field up and down, respectively.
With the experimentally determined value $k=1.90\pm0.25\times10^{-26}%
\ e$-cm/ppm, Eq. (\ref{6a}) indicates an Earth rotation EDM value of
$d_{n\oplus}=(2.57\pm0.34)\times10^{-26}\ e\ \mathrm{cm}.$
which should have been experimentally observed with the analysis procedure and
therefore serves as a sort of ``blind analysis" parameter.

Many of the systematic effects in Table I of \cite{1}
that contribute $d_{n}^{\prime}$ result from shifts in
$R_{a0}$; the Earth rotation effect, $d_{n\oplus}$, is about five times
larger than the largest entry in that table. Including $d_{n\oplus}$ in the table
and following the correction procedure of \cite{1} results in 
$d_n=-(2.4\pm 1.5\ ({\rm stat})\pm 0.5\ ({\rm syst}))\times
10^{-26}\ e\ {\rm cm}$.
Therefore the systematic uncertainty and 90\% confidence limit
appear as underestimated in \cite{1}.

\end{document}